\documentclass[a4paper]{article}

\usepackage{INTERSPEECH2021}
\usepackage{float}
\usepackage{multirow}
\usepackage{soul}

\title{Lightweight Dual-channel Target Speaker Separation for Mobile Voice Communication}
\name{Yuanyuan Bao$^{1,\text{*}}$$\thanks{\hspace{-1mm}* Equal Contribution}$, Yanze Xu$^{1,\text{*}}$, Na Xu$^2$, Wenjing Yang$^2$, Hongfeng Li$^2$, Shicong Li$^2$, Yongtao Jia$^2$, \\
Fei Xiang$^2$, Jincheng He$^1$, Ming Li$^{1,\thanks{corresponding author: Ming Li, ming.li369@dukekunshan.edu.cn}}$}
\address{
    $^1$Data Science Research Center, Duke Kunshan University, Kunshan, China\\
    $^2$Xiaomi, Beijing, China}
\email{ming.li369@dukekunshan.edu.cn}

\begin{document}

\maketitle
\begin{abstract}
        Nowadays, there is a strong need to deploy the target speaker separation (TSS) model on mobile devices with a limitation of the model size and computational complexity. To better perform TSS for mobile voice communication, we first make a dual-channel dataset based on a specific scenario, LibriPhone. Specifically, to better mimic the real-case scenario, instead of simulating from the single-channel dataset, LibriPhone is made by simultaneously replaying pairs of utterances from LibriSpeech by two professional artificial heads and recording by two built-in microphones of the mobile.
        Then, we propose a lightweight time-frequency domain separation model, LSTMFormer, which is based on the LSTM framework with source-to-noise ratio (SI-SNR) loss.
        For the experiments on LibriPhone, we explore the dual-channel LSTMFormer model and a single-channel version by a random single channel of LibriPhone.
        Experimental result shows that the dual-channel LSTMFormer outperforms the single-channel LSTMFormer with relative 25\% improvement. This work provides a feasible solution for the TSS task on the mobile devices, playing back and recording multiple data sources in real application scenarios for getting dual-channel real data can assist the lightweight model to achieve higher performance.

    \end{abstract}

    \section{Introduction}
    Cocktail party effect \cite{C1}, or cocktail party problem, refers to the phenomenon that human beings can easily focus on the interested voice with background noises and multiple speakers talking around.

    As an effective solution to solve the above problem, speech separation was proposed and has been studied for decades. It decomposes the given mixed utterance into individual sources. 
    Deep clustering (DPCL) \cite{C4, dpcl}, deep attractor network (DANet) \cite{C5, DAnet} and permutation invariant training (PIT) \cite{pit, C6} are very representative works.
    Breakthrough is made by time-domain audio source separation (Tas-Net) \cite{C7, tas2, tas3} and Dual-path RNN (DPRNN) \cite{C8}, which introduce the source-to-noise ratio (SI-SNR) loss for separation and take the phase information into account.

    Despite the great progress made, the aforementioned methods all require the number of speakers to be known in advance in the training or inference stage. It limits the usage of speech separation in real-world applications, and in reality, only the target speaker's voice is needed in most cases, such as talking on the phone and waking up the voice assistant. Thus a new technique, namely target speaker separation (TSS) \cite{C9, C11} or speaker extraction \cite{C10, spex, spex+, spex++}, is brought up in recent years. It could separate the target speaker's voice from the mixed utterance when the reference utterance of the target speaker is available.

    Studies on this technique have achieved impressive and heuristic performance. Single-channel work like the time-frequency domain model VoiceFilter \cite{C11} and time-domain model SpEx+ \cite{spex+} use the speaker embedding extracted from the reference utterance to help the model learn whom to separate. Multi-channel work like \cite{C14} makes use of the location information of the target speaker to get the utterance wanted. In the multi-modal domain, researchers incorporate visual information into the speech separation system for better separation performance \cite{C15}.

    The mentioned studies do achieve state-of-the-art results on public speech separation datasets. However, there are still some limitations when it comes to real-world applications. First, due to the restriction of computational resources on some platforms, e.\ g. mobile devices, cell phone, the model size, and computing complexity are often constrained strictly.
    And the real constriction is on the separation module rather than the speaker verification module. Specifically, the target speaker's utterance is usually enrolled in advance, thus the speaker embedding could be extracted and saved beforehand.
    So only the parameters and multiply-accumulate operations (MACs) cost of the separation module are considered and reported. However,  to improve the separation accuracy, speech separation modules always have a large number of parameters and deep network design.
    For instance, the Long short-term memory (LSTM) based models usually contain hundreds even thousands of hidden units, which results in large model size and complexity. The Atss-Net \cite{C16} do decrease the number of parameters, while the computational complexity is still very high. Another issue is the need for causal and real time processing which many attention based models can not satisfy.
    The third difficulty for the application on mobile is that there is a domain mismatch between the real scenario of mobile voice communication and that of the simulated available public corpus. Moreover, the dual-channel data used for dual-channel separation is always simulated from the single-channel public corpus \cite{dual}.
    The domain gaps caused by both the simulation and the difference of the scenarios may lead to the degradation of the real-case inference performance.

    To overcome the aforementioned problems for the application on mobile devices, we first record a dual-channel dataset called LibriPhone to mimic the mobile voice communication scenario. Specifically, instead of simulating mixed data by direct addition, LibriPhone is collected in a relatively real setting in which two random utterances from LibriSpeech are simultaneously played by two professional artificial heads and recored by two built-in microphones in a cell phone.

    Based on our previous work in Atss-Net \cite{C16}, we design a lightweight time-frequency domain separation model and share the same speaker verification module with Atss-Net. The reason for choosing the time-frequency domain model is that it has a smaller input dimension than the time-domain model. Inspired by the effort made in lightweight model development in speech separation domain \cite{C17, C18} and the introduction of the SI-SNR loss \cite{C7}, we design a lightweight target speaker separation model, LSTMFormer, based on the LSTM network and incorporate SI-SNR loss for better performance. Nonetheless, re-recording will inevitably introduce delays between the original clean utterance sent to artificial mouth and the one received by the microphones caused by the recording equipment. This may cause the non-convergence problem for training. The reason is that the computation of the SI-SNR loss is not robust to the time delay. Therefore, we perform generalized cross-correlation \cite{gcc} based alignment as a training strategy to address this problem.

    For the experiments on LibriPhone, we explore both the dual-channel and single-channel versions of LSTMFormer. Specifically, both single- and dual-channel versions work well on LibriPhone. Moreover, we discover that the dual-channel model outperforms the single-channel one significantly when they have similar model size. It shows that the dual-channel data is easier for the lightweight model to learn and to achieve better performance.

    \section{DataSet Description \label{sec:2}}


    The LibriPhone is a dual-channel dataset re-recorded from LibriSpeech for mobile voice communication scenario. There are 25 hours 16 kHz sampled mixed utterances in total and split with 15 hours for training, 5 hours for validating, and 5 hours for testing. The subsets for training and validating is made from train-clean-100, the validation subset is made from dev-clean. In each subset, we respectively playback and record the mixed utterance and target utterance. Specifically, for each mixed utterance, we select a pair of utterances, the target speaker utterance, and the other speaker utterance, with non-repetition of the target speaker identity. And then, two artificial heads separately replay two clean utterances simultaneously. A cell phone with two microphones is used to record the mixed sound. Likewise, we also record the target speaker utterances as the ground truth for supervised learning with the similar setup but only one artificial head replaying the target speaker utterance.

\begin{figure}[htbp]
        \centering
        \includegraphics[width=0.8\linewidth]{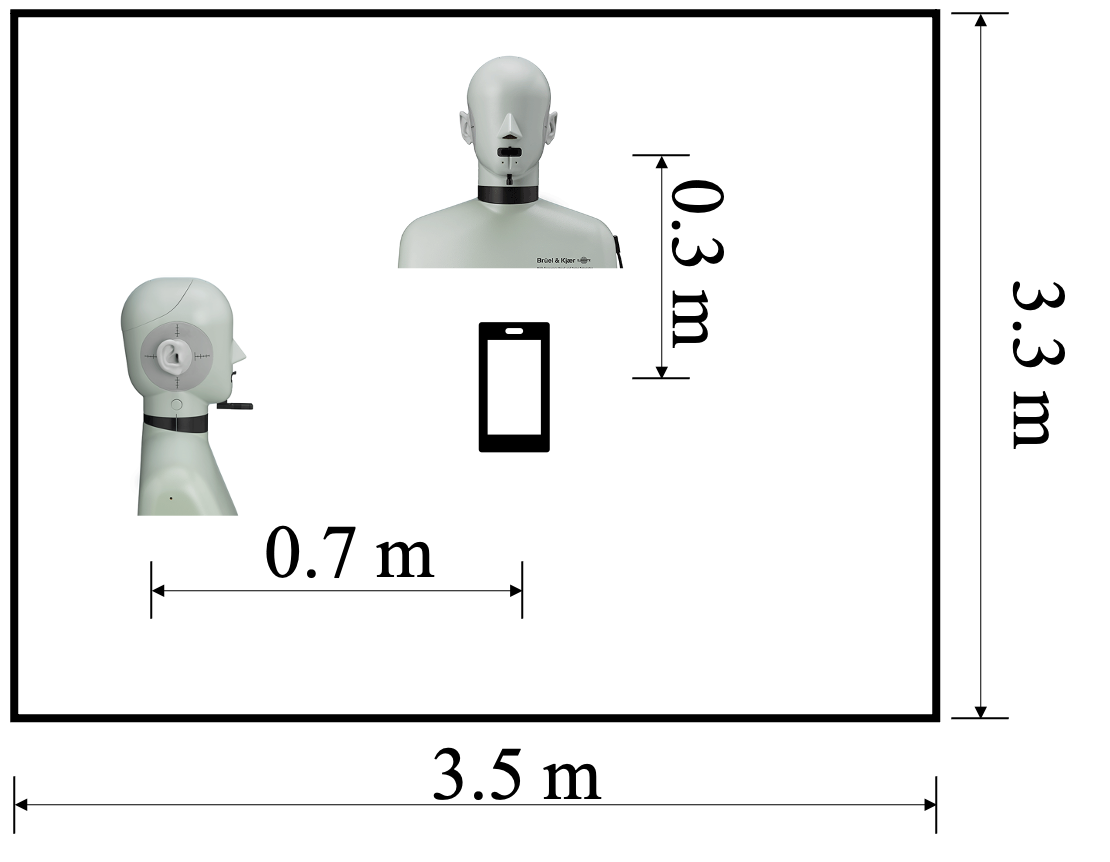}
        \caption{The recording setup of our proposed LibriPhone database.}
        \label{room}
    \end{figure}

    The detailed setup is shown in Figure~\ref{room}. The recording equipment is an Android phone placed in the center of the room. The room is 3.3 meters wide and 3.5 meters long with a height of 2.3 meters. Two artificial heads are used to mimic people talking. The one closer to the phone is the target speaker, the other is the interference. The specific locations of both artificial heads are illustrated in Figure~\ref{room}. The mouths of both artificial heads are placed at 1.5m height, the screen of the phone is horizontal and orients towards the target speaker. By measuring the acoustic parameters of the recording room, the background noise level is 25 dBA, the reverberation time $T_{60}$ is 0.45s to 0.55s.

    \section{Model Designing}
    Our TSS framework shares the same speaker verification module with our previous work in \cite{C16}. Due to the high computational complexity of the Transformer-encoder-based separation module, when designing LSTMFormer, we replace the self-attention in the Transformer encoder with a single direction LSTM and keep the idea of skip connection and layernorm.

    In this work, we want to explore whether the real-collected dual-channel data in our simulated acoustic studio can achieve superior separation perforamance. Hence, we first designed a dual-channel version LSTMFormer on the LibriPhone. And then, for comparison, we also design a single-channel version LSTMFormer trained by using the random single channel of LibriPhone. The two versions of the LSTMFormer are respectively introduced in section~\ref{sec:dual model} and section~\ref{sec:single model}.

    \subsection{Dual-channel LSTMFormer \label{sec:dual model}}
\begin{figure}[H]
        \centering
        \includegraphics[width=\linewidth]{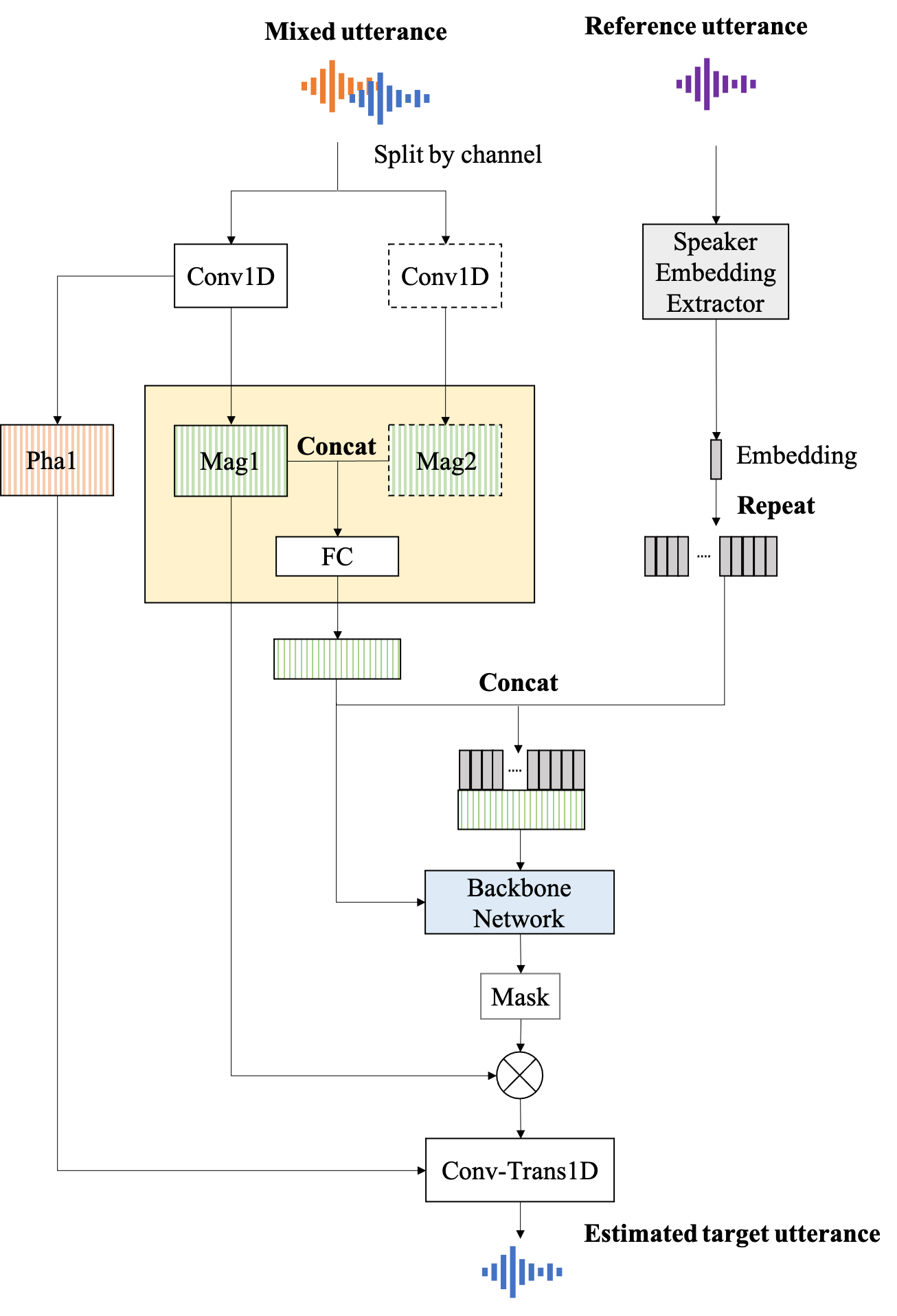}
        \caption{
            Solid part + Dashed part: Model achitecture of Dual-channel LSTMFormer. Solid part only: Model achitecture of Single-channel LSTMFormer. Mag1 and Mag2 represent the magnitude spectrogram of the split two channels of mixed utterance. Pha1 is the phase spectrogram of the first channel used to reconstruct the target utterance. The box in yellow is used to compress Mag1 and Mag2.}
        \label{pic1}
    \end{figure}

    As shown in Figure~\ref{pic1}, an STFT Conv1D layer is used to transform the mixed dual-channel waveform $\mathbf{x}\in \mathbb{R}^{2\times L}$ into magnitude and phase spectrogram, where $L$ is the length of sample points, and $\mathbf{x}$ is a linear combination of $C$ sources $s_{1}(t), \ldots, s_{c}(t)$. The mixed dual-channel utterance is first split into two channels and passed through the Conv1D layer separately.

    \begin{equation}
        M_{i}, P_{i}=\operatorname{Conv1D}\left(channel_{i}\right)  \quad i=1,2
    \end{equation}
    where $M_i$, $P_i$ $\in \mathbb R^{T\times F}$ stand for magnitude and phase spectrogram respectively. $T$ and $F$ are the dimension of frames and spectrogram bin axes.

    As the yellow block in Figure~\ref{pic1} shows, the two magnitude spectrogram are firstly concatenated along the feature dimension and then compressed using fully-connected layer for feature extraction.

\begin{figure}[ht]
        \centering
        \includegraphics[width=\linewidth]{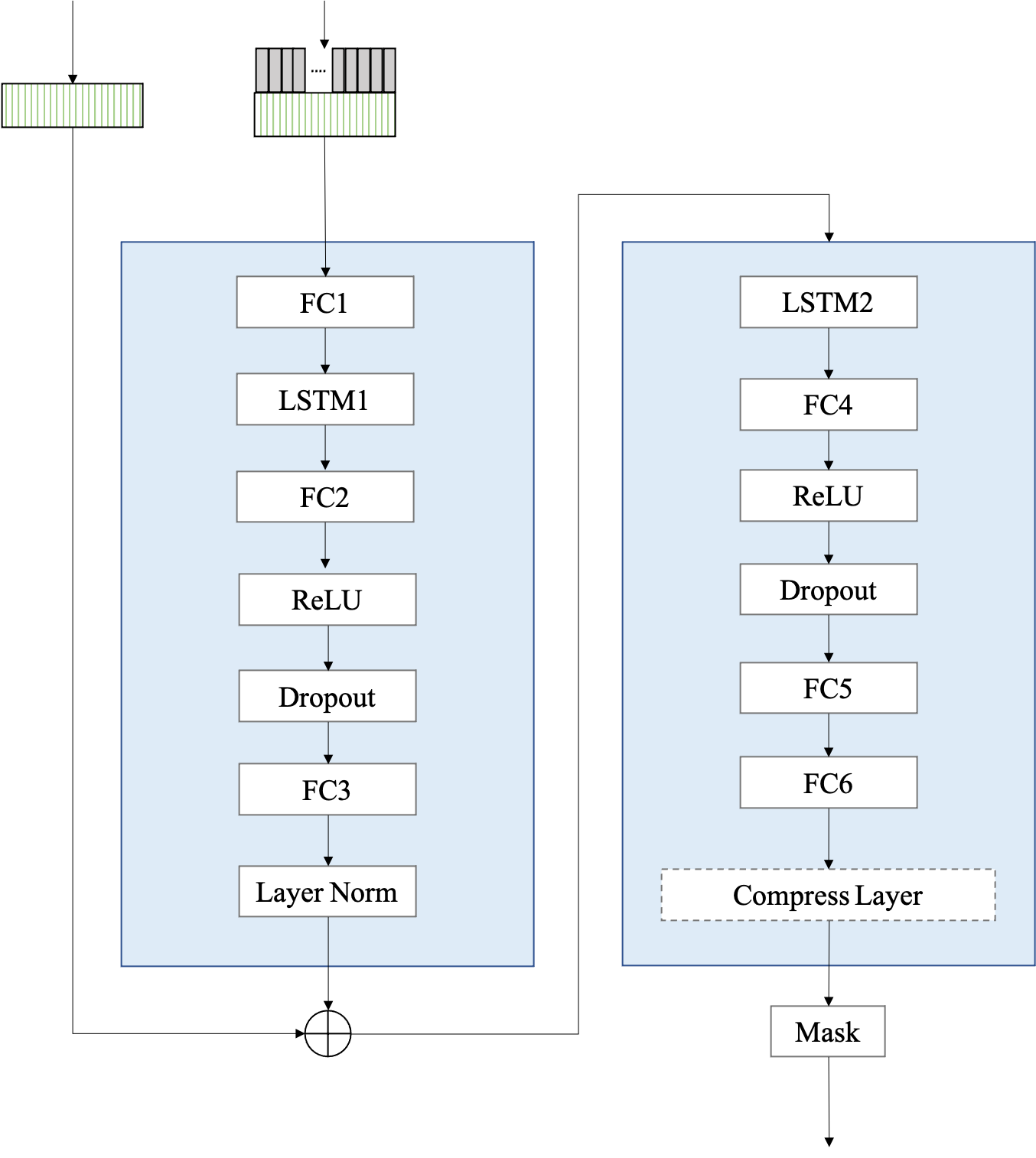}
        \caption{The backbone network of the LSTMFormer model.}
        \label{pic2}
    \end{figure}

    The speaker embedding is repeated for $T$ times for the ease of concatenation. The extracted feature map and the repeated speaker embedding are then concatenated along the feature dimension and fed into the following LSTMFormer. The backbone network of our proposed LSTMFormer model is shown in Figure~\ref{pic2}. Motivated by the idea of residual learning \cite{he2016deep}, there is a skip connection between the compressed magnitude spectrogram and the Layer Norm output. This design makes our network better remember the information of compressed magnitude spectrogram.The fully-connected layers (FC) are used for deeper feature extraction as well as the flexible dimension changes between different modules of the whole model. An extra FC layer in the end serves as the compress layer to reduce the dimension to match with the time-frequency mask for a single reference channel. Specifically, we choose the first channel as the reference channel. The estimated magnitude spectrogram is calculated by performing the element-wise product between the mixed magnitude spectrogram of the first channel $M_1$ and the estimated mask $R\in \mathbb{R}^{T\times F}$. With the estimated magnitude spectrogram and phase spectrogram of the first channel, the iSTFT Conv-Trans1D outputs the estimated target speaker's utterance $\hat{\mathbf{s}}\in \mathbb{R}^{1\times L}$.
    

    \begin{equation}
        \hat{\mathbf{s}}=\operatorname{Conv-Trans1D}\left(M_1\odot R, P_1\right)
    \end{equation}
    where $\odot$ denotes the element-wise product operation.

    As the input and output of this whole separation module are both waveform, we choose scale-invariant source-to-noise ratio (SI-SNR) as our training target, to directly optimize the separation performance \cite{C7}. The SI-SNR is defined as follows:\\
    \begin{equation}
        \mathbf{s}_{\text {target }}=\frac{\langle\hat{\mathbf{s}}, \mathbf{s}\rangle \mathbf{s}}{\|\mathbf{s}\|^{2}}
    \end{equation}
    \begin{equation}
        \mathbf{e}_{\text {noise }}=\hat{\mathbf{s}}-\mathbf{s}
    \end{equation}

    \begin{equation}
        \text {SI-SNR}:=10 \log _{10} \frac{\left\|\mathbf{s}_{\text {target }}\right\|^{2}}{\left\|\mathbf{e}_{\text {noise }}\right\|^{2}}
    \end{equation}
    where $\mathbf{s}\in \mathbb{R}^{1\times L}$ is the ground truth target speaker's utterance.

    \begin{table*}[th]
    \scriptsize
        \renewcommand\arraystretch{1.5}
        \caption{The model configurations of single- and dual- channel LSTMFormer models. Single-channel (half) and single-channel (equal) refer to half and equal model parameters in backbone network of the dual-channel model, respectively.}
        \vskip5pt
        \label{tab:setup}
        \centering
        \begin{tabular}{c|c|c|c|c}
            \toprule[1pt]
            \multicolumn{2}{c|}{\textbf{Layer}} & \textbf{Dual-channel} & \textbf{Single-channel (half)}
            & \textbf{Single-channel (equal)}
            \\ \hline
            \multicolumn{2}{c|}{Conv1D}         & $[514\times 400, 2]$ & $[514\times 400, 1]$ & $[514\times 400, 1]$ \\ \hline
            \multicolumn{2}{c|}{FC}             & $[1\times 1, 256]$   & $[1\times 1, 128]$   & $[1\times 1, 256]$   \\ \hline
            \multicolumn{2}{c|}{FC1}            & $[1\times 1, 256]$   & $[1\times 1, 128]$   & $[1\times 1, 256]$   \\ \hline
            \multicolumn{2}{c|}{LSTM1} & $hidden\_state_{num} = 256$ & $hidden\_state_{num} = 128$
            & $hidden\_state_{num} = 256$
            \\ \hline
            \multicolumn{2}{c|}{FC2}            & $[1\times 1, 200]$   & $[1\times 1, 100]$   & $[1\times 1, 200]$   \\ \hline
            \multicolumn{2}{c|}{FC3}            & $[1\times 1, 256]$   & $[1\times 1, 128]$   & $[1\times 1, 256]$   \\ \hline
            \multicolumn{2}{c|}{LSTM2} & $hidden\_state_{num} = 256$ & $hidden\_state_{num} = 128$
            & $hidden\_state_{num} = 256$
            \\ \hline
            \multicolumn{2}{c|}{FC4}            & $[1\times 1, 180]$   & $[1\times 1, 90]$    & $[1\times 1, 180]$   \\ \hline
            \multicolumn{2}{c|}{FC5}            & $[1\times 1, 256]$   & $[1\times 1, 128]$   & $[1\times 1, 256]$   \\ \hline
            \multicolumn{2}{c|}{FC6}            & $[1\times 1, 514]$   & $[1\times 1, 257]$   & $[1\times 1, 514]$   \\ \hline
            \multicolumn{2}{c|}{Compress Layer} & $[1\times 1, 257]$   & N/A                  & N/A                  \\ \hline
            \multicolumn{2}{c|}{Conv-Trans1D}   & $[514\times 400, 1]$ & $[514\times 400, 1]$ & $[514\times 400, 1]$ \\
            \bottomrule[1pt]
        \end{tabular}
    \end{table*}

    \subsection{Single-channel LSTMFormer\label{sec:single model}}
    For the single-channel version, as shown in Figure~\ref{pic1}, removing modules with the dashed line, including one STFT Conv1D layer and one compress layer of the dual-channel version. There are two kinds of single-channel LSTMFormer with the same architectures but different model size. One of the model size of single-channel LSTMFormer is decreased to half by halving the hidden numbers of the LSTM and the output dimension of the FC layer. Another one's model size is close to the dual-channel LSTMFormer.

    \section{Experiment Setup}

    \subsection{Model Training}
    All of our systems are implemented using PyTorch \cite{pytorch}. Adam serves as the optimizer during training \cite{adam} with an initial learning rate of 0.0001. All models are trained with the batch size of 64 and stopped till the model performance didn't improve on the validation set any longer. The model configurations of all the models related to LSTMFormer are shown in table1. The single-channel model shares the same architecture with the dual-channel one using half of the  parameters in the dual-channel LSTMFormer module. To find out the impact of increasing model parameters and validate the advantage of dual-channel complementary information in the TSS task, we trained another single-channel model with as much parameters as the dual-channel model. During the training of the single-channel model, we randomly chose one channel of the mixed utterance. Signal-to-distortion ratio (SDR) is used as the evaluation metric for the TSS task.


    \subsection{Alignment}
    The unstable time delay between the target and mixed utterances in our LibriPhone dataset is found to be a big barrier during the training stage. To overcome it, we compute the time delay of each target and mixed utterance pairs based on the generalized cross-correlation method implemented in Matlab\footnote{https://www.mathworks.com/help/phased/ref/gccphat.html} for better alignment.

    \section{Results and Discussion}

    For the performance of different models on the LibriPhone dataset, Table 2 shows the before-separated, after-separated SDR values and the relative improvement, the information on the model size and computational complexity are also reported.

    From Table 2, we can see that both dual- and single- LSTMFormer achieve an over 4.8dB SDR improvement, which means alignment can well address the problem that the SI-SNR loss can not converge because of the delay introduced by the re-recording. We can also find that the MACs cost of any model with 1 second input speech is very low, which requires less computation resources of the  application platforms.

    Specifically, the improvement of single-channel (half) with 36.32M MACs is 4.85dB, and the improvement of single-channel(equal) with 128.45M MACs is 5.12dB. The increase of the model size for single-channel LSTMFormer only improves 0.27dB SDR. However, the dual-channel LSTMFormer with 154.29M MACs achieves a 7.71dB improvement, a relative 25\% improvement compared with the single-channel model. It shows that dual-channel data can help the lightweight model LSTMFormer to  achieve better performance. This motivates us to explore whether the multi-channel data can improve the lightweight model further than the dual-channel data. In a similar way we performed, we can re-record multi-channel data to mimic more complex application scenarios. Without explicitly using the location and direction information, our work provides a feasible solution to the dual-channel TSS task.

    \begin{table}[th]
    \scriptsize
        \renewcommand\arraystretch{1.5}
        \caption{Performance of different models on the LibriPhone test set. The model size and MACs cost on 1-second inputs of different models are also reported.}
        \centering
        \resizebox{\columnwidth}{!}{
        \begin{tabular}{cccccc}
            \toprule[1pt]
            \multirow{2}*{\textbf{Model}} &\multirow{2}*{\textbf{Params}} &\multirow{2}*{\textbf{MACs}} &\multicolumn{3}{c}{\textbf{Mean SDR}} \\ \cline{4-6}
            &  &      & Before & After & Improved \\ \hline
            Single-channel (half) & 0.37M           & 36.32M & 3.12   & 7.97  & 4.85     \\
            Single-channel (equal) & 1.31M           & 128.45M & 3.22   & 8.34  & 5.12     \\ \hline
            Dual-channel         & 1.57M           & 154.29M  & 2.69   & 10.40 & 7.71     \\
            \bottomrule[1pt]
        \end{tabular}}
        \label{tab:my_label}
    \end{table}



    \section{Conclusions}

    In this paper, to reduce the domain gap of dual-channel TSS systems for mobile voice communication, we first design and collect a dual-channel dataset, LibriPhone. Specifically, to better mimic the real application scenario, instead of directly mixing the single-channel dataset,  LibriPhone is made by simultaneously replaying pairs of utterances from LibriSpeech by two artificial heads and recording by two microphones of the mobile device. And then, we propose a lightweight time-frequency domain separation model, LSTMFormer for the experiments on LibriPhone. We explore both the dual-channel and the single-channel LSTMFormer model. Experimental result shows that the dual-channel model outperforms the single-channel significantly when they have the same model size, and when the former is twice bigger than the latter, showing that the dual-channel data are complementary and easier for the lightweight model to learn. Our proposed lightweight model, and corresponding training strategies demonstrate the promising prospects of deep-learning-based TSS models being applied on real-world low-resource platforms.

   
    \bibliographystyle{IEEEtran}

    \bibliography{template}

\begin{thebibliography}{10}
\providecommand{\url}[1]{#1}
\csname url@samestyle\endcsname
\providecommand{\newblock}{\relax}
\providecommand{\bibinfo}[2]{#2}
\providecommand{\BIBentrySTDinterwordspacing}{\spaceskip=0pt\relax}
\providecommand{\BIBentryALTinterwordstretchfactor}{4}
\providecommand{\BIBentryALTinterwordspacing}{\spaceskip=\fontdimen2\font plus
\BIBentryALTinterwordstretchfactor\fontdimen3\font minus
  \fontdimen4\font\relax}
\providecommand{\BIBforeignlanguage}[2]{{%
\expandafter\ifx\csname l@#1\endcsname\relax
\typeout{** WARNING: IEEEtran.bst: No hyphenation pattern has been}%
\typeout{** loaded for the language `#1'. Using the pattern for}%
\typeout{** the default language instead.}%
\else
\language=\csname l@#1\endcsname
\fi
#2}}
\providecommand{\BIBdecl}{\relax}
\BIBdecl

\bibitem{C1}
E.~C. Cherry, ``Some experiments on the recognition of speech, with one and
  with two ears,'' \emph{The Journal of the acoustical society of America},
  vol.~25, no.~5, pp. 975--979, 1953.

\bibitem{C4}
J.~R. Hershey, Z.~Chen, J.~Le~Roux, and S.~Watanabe, ``Deep clustering:
  Discriminative embeddings for segmentation and separation,'' in
  \emph{ICASSP}.\hskip 1em plus 0.5em minus 0.4em\relax IEEE, 2016, pp. 31--35.

\bibitem{dpcl}
Y.~Isik, J.~Le~Roux, Z.~Chen, S.~Watanabe, and J.~R. Hershey, ``Single-channel
  multi-speaker separation using deep clustering,'' in
  \emph{Interspeech}.\hskip 1em plus 0.5em minus 0.4em\relax IEEE, 2016, pp.
  545--549.

\bibitem{C5}
Z.~Chen, Y.~Luo, and N.~Mesgarani, ``Deep attractor network for
  single-microphone speaker separation,'' in \emph{ICASSP}.\hskip 1em plus
  0.5em minus 0.4em\relax IEEE, 2017, pp. 246--250.

\bibitem{DAnet}
Y.~Luo, Z.~Chen, and N.~Mesgarani, ``Speaker-independent speech separation with
  deep attractor network,'' \emph{IEEE/ACM Transactions on Audio, Speech, and
  Language Processing}, vol.~26, no.~4, pp. 787--796, 2018.

\bibitem{pit}
D.~Yu, M.~Kolb{\ae}k, Z.-H. Tan, and J.~Jensen, ``Permutation invariant
  training of deep models for speaker-independent multi-talker speech
  separation,'' in \emph{ICASSP}.\hskip 1em plus 0.5em minus 0.4em\relax IEEE,
  2017, pp. 241--245.

\bibitem{C6}
M.~Kolb{\ae}k, D.~Yu, Z.-H. Tan, and J.~Jensen, ``Multitalker speech separation
  with utterance-level permutation invariant training of deep recurrent neural
  networks,'' \emph{IEEE/ACM Transactions on Audio, Speech, and Language
  Processing}, vol.~25, no.~10, pp. 1901--1913, 2017.

\bibitem{C7}
Y.~Luo and N.~Mesgarani, ``Tasnet: time-domain audio separation network for
  real-time, single-channel speech separation,'' in \emph{ICASSP}.\hskip 1em
  plus 0.5em minus 0.4em\relax IEEE, 2018, pp. 696--700.

\bibitem{tas2}
Y.~Luo and N.~Mesgarani, ``Real-time single-channel dereverberation and
  separation with time-domain audio separation network.'' in
  \emph{Interspeech}.\hskip 1em plus 0.5em minus 0.4em\relax IEEE, 2018, pp.
  342--346.

\bibitem{tas3}
Y.~Luo and N.~Mesgarani, ``Conv-tasnet: Surpassing ideal time--frequency
  magnitude masking for speech separation,'' \emph{IEEE/ACM transactions on
  audio, speech, and language processing}, vol.~27, no.~8, pp. 1256--1266,
  2019.

\bibitem{C8}
Y.~Luo, Z.~Chen, and T.~Yoshioka, ``Dual-path rnn: efficient long sequence
  modeling for time-domain single-channel speech separation,'' in
  \emph{ICASSP}.\hskip 1em plus 0.5em minus 0.4em\relax IEEE, 2020, pp. 46--50.

\bibitem{C9}
J.~Du, Y.~Tu, Y.~Xu, L.~Dai, and C.~H. Lee, ``Speech separation of a target
  speaker based on deep neural networks,'' in \emph{ICSP}.\hskip 1em plus 0.5em
  minus 0.4em\relax IEEE, 2014, pp. 473--477.

\bibitem{C11}
Q.~Wang, H.~Muckenhirn, K.~Wilson, P.~Sridhar, Z.~Wu, J.~R. Hershey, R.~A.
  Saurous, R.~J. Weiss, Y.~Jia, and I.~L. Moreno, ``Voicefilter: Targeted voice
  separation by speaker-conditioned spectrogram masking,'' in
  \emph{Interspeech}.\hskip 1em plus 0.5em minus 0.4em\relax IEEE, 2019, pp.
  2728--2732.

\bibitem{C10}
C.~Xu, W.~Rao, E.~S. Chng, and H.~Li, ``Time-domain speaker extraction
  network,'' in \emph{ASRU}.\hskip 1em plus 0.5em minus 0.4em\relax IEEE, 2019,
  pp. 327--334.

\bibitem{spex}
C.~Xu, W.~Rao, E.~S. Chng, and H.~Li, ``Spex: Multi-scale time domain speaker
  extraction network,'' \emph{IEEE/ACM Transactions on Audio, Speech, and
  Language Processing}, vol.~28, pp. 1370--1384, 2020.

\bibitem{spex+}
M.~Ge, C.~Xu, L.~Wang, E.~S. Chng, J.~Dang, and H.~Li, ``Spex+: A complete time
  domain speaker extraction network,'' in \emph{Proc.Interspeech},
  vol.~28.\hskip 1em plus 0.5em minus 0.4em\relax IEEE, 2020, pp. 1406--1410.

\bibitem{spex++}
M.~Ge, C.~Xu, L.~Wang, E.~S. Chng, J.~Dang, and H.~Li, ``Multi-stage speaker
  extraction with utterance and frame-level reference signals,'' in
  \emph{ICASSP}.\hskip 1em plus 0.5em minus 0.4em\relax IEEE, 2021, pp.
  6109--6113.

\bibitem{C14}
G.~Li, S.~Liang, S.~Nie, W.~Liu, M.~Yu, L.~Chen, S.~Peng, and C.~Li,
  ``Direction-aware speaker beam for multi-channel speaker extraction,'' in
  \emph{Interspeech}.\hskip 1em plus 0.5em minus 0.4em\relax IEEE, 2019, pp.
  2713--2717.

\bibitem{C15}
R.~Gu, S.~Zhang, Y.~Xu, L.~Chen, Y.~Zou, and D.~Yu, ``Multi-modal multi-channel
  target speech separation,'' \emph{IEEE Journal of Selected Topics in Signal
  Processing}, vol.~14, no.~3, pp. 530--541, 2020.

\bibitem{C16}
T.~Li, Q.~Lin, Y.~Bao, and M.~Li, ``Atss-net: Target speaker separation via
  attention-based neural network,'' in \emph{Proc. Interspeech}, 2020, pp.
  1411--1415.

\bibitem{dual}
C.~Li, J.~Xu, N.~Mesgarani, and B.~Xu, ``Speaker and direction inferred
  dual-channel speech separation,'' in \emph{ICASSP}.\hskip 1em plus 0.5em
  minus 0.4em\relax IEEE, 2021, pp. 5779--5783.

\bibitem{C17}
Y.~Luo, C.~Han, and N.~Mesgarani, ``Ultra-lightweight speech separation via
  group communication,'' in \emph{ICASSP}.\hskip 1em plus 0.5em minus
  0.4em\relax IEEE, 2021, pp. 16--20.

\bibitem{C18}
E.~Tzinis, Z.~Wang, and P.~Smaragdis, ``Sudo rm-rf: Efficient networks for
  universal audio source separation,'' in \emph{MLSP}.\hskip 1em plus 0.5em
  minus 0.4em\relax IEEE, 2020, pp. 1--6.

\bibitem{gcc}
C.~Knapp and G.~Carter, ``The generalized correlation method for estimation of
  time delay,'' \emph{IEEE transactions on acoustics, speech, and signal
  processing}, vol.~24, no.~4, pp. 320--327, 1976.

\bibitem{he2016deep}
K.~He, X.~Zhang, S.~Ren, and J.~Sun, ``Deep residual learning for image
  recognition,'' in \emph{Proceedings of the IEEE conference on computer vision
  and pattern recognition}, 2016, pp. 770--778.

\bibitem{pytorch}
A.~Paszke, S.~Gross, F.~Massa, A.~Lerer, J.~Bradbury, G.~Chanan, T.~Killeen,
  Z.~Lin, N.~Gimelshein, L.~Antiga \emph{et~al.}, ``Pytorch: An imperative
  style, high-performance deep learning library,'' \emph{arXiv preprint
  arXiv:1912.01703}, 2019.

\bibitem{adam}
D.~P. Kingma and J.~Ba, ``Adam: A method for stochastic optimization,''
  \emph{arXiv preprint arXiv:1412.6980}, 2014.

\end{thebibliography}


\end{document}